\documentclass[prd, twocolumn, nofootinbib, floatfix]{revtex4-2} 

\usepackage{amsmath}
\usepackage{graphicx}
\usepackage{dcolumn}
\usepackage{bm}
\usepackage{epsfig}
\usepackage{amssymb,latexsym,mathrsfs,empheq}
\usepackage{graphicx}
\usepackage{color}
\usepackage{xcolor} 
\usepackage{mathtools} 
\usepackage{hyperref}

\hypersetup{
    colorlinks=true,
    linkcolor=red,
    citecolor=blue,
} 

\newcommand{\be}{\begin{equation}}
\newcommand{\ee}{\end{equation}}
\newcommand{\bs}{\begin{split}} 
\newcommand{\bea}{\begin{eqnarray}}
\newcommand{\eea}{\end{eqnarray}}

\newcommand{\om}{\Omega_m} 
\newcommand{\wde}{w_{\rm de}}
\newcommand{\rde}{\rho_{\rm de}} 

\newcommand{\gm}{G_{\rm matter}} 
\newcommand{\gl}{G_{\rm light}}

\begin{document}

\title{Benchmarks of Dark Energy} 

\author{Eric V.\ Linder} 
\affiliation{
Berkeley Center for Cosmological Physics \& Berkeley Lab, 
University of California, Berkeley, CA 94720, USA
} 

\begin{abstract}
Dark energy is a premier mystery of physics, both theoretical and
experimental. As we look to develop plans for high energy physics
over the next decade, within a two decade view, we consider benchmarks
for revealing the nature of dark energy. We conclude, based on
fundamental physical principles detailed below,  that understanding
will come from experiments reaching key benchmarks:
\begin{itemize}
\item $\sigma(w_a)<2.5\sigma(w_0)$
\item $\sigma(w_0)<0.02$
\item $\sigma(\rho_{\rm de}/\rho_{\rm crit})<(1/3)\rho_\Lambda/\rho_{\rm crit}$
for all redshifts $z<5$
\end{itemize}
where the dark energy equation of state $w(a)=w_0+w_a(1-a)$. Beyond
the cosmic expansion history we also discuss benchmarks for the cosmic
growth history appropriate for testing classes of gravity theories. 
All benchmarks can be achieved by a robust Stage 5 program, using 
extensions of existing probes plus the highly complementary, 
novel probe of cosmic redshift drift. 
\end{abstract} 

\date{\today} 

\maketitle

\section{Introduction} 

Cosmic acceleration, in its late time instantiation known as dark
energy, was highlighted at Snowmass 2001 as a premier problem, with
quotes such as ``the problem of dark energy will be a roadblock on
our path to a comprehensive fundamental physical theory'', ``the
biggest and most profound gap in our current understanding of the physical
world'', ``it is vitally important to know if the cosmological `constant'
is truly constant'' \cite{resbkde}. These comments are as true now.

In this article, as succinctly as possible, we present the outstanding
questions about the nature of dark energy, and the benchmarks that will
still need to be met after the current (``Stage 4'') dark energy
experiments complete. It is challenging to set benchmarks -- revelatory
measurement levels -- for a physical phenomenon that is not understood.
To do so we rely on basic physical principles and observations, e.g.\
our universe is many e-folds of expansion old, it had a long period of
matter domination before the current acceleration, etc.\ plus an
appeal to avoidance of fine tuning to whatever extent is viable.

Another guiding principle is that for phenomenology we choose as few
parameters as can robustly describe the impact of theory on observations,
and for theory we choose as few parameters as can showcase major
characteristics of classes of theory (in particular for modified
gravity). That is, we are looking for an alert, a signal, a direction,
and want the highest significance indication as to   signatures of dark
energy. This minimalist approach likely shortchanges the rich
tapestry of the dark sector, but has precedent in that most of late
time cosmic expansion and growth for example can neglect the intricacies of the
Standard Model of particle physics and simply investigate the matter density.
Once observations give key hints based on the benchmarks, then deeper
investigation will burst forth. 

Section~\ref{sec:exp} examines the cosmic expansion history over
the last 10 billion years, while Section~\ref{sec:z25} motivates
the investigation of dark energy before it came to dominance. 
Section~\ref{sec:gro} explores the cosmic growth history, the
importance of testing consistency with the expansion history, and
how modified gravity manifests. Section~\ref{sec:exper} provides
some thoughts on how to focus experiments (``Stage 5'') to achieve
the benchmarks.

\section{Cosmic Acceleration} \label{sec:exp}

Cosmic expansion is determined by the Friedmann equations, giving
the expansion rate or Hubble parameter $H=\dot a/a$ in terms of the
effective energy-momenta of the components. For a homogeneous and
isotropic universe these resolve to the energy density and pressure
of each component, and the ratio $w_i(a)=P_i(a)/\rho_i(a)$ is known
as the equation of state parameter (EOS). This holds even outside
general relativity by regarding the energy density and pressure of 
an effective component. The late time matter component is well treated
as pressureless, $w_m=0$. A cosmological constant $\Lambda$, with time independent
energy density, has $P_\Lambda=-\rho_\Lambda$ and so $w_\Lambda=-1$.

A central question is what is the EOS of dark energy -- is it a
cosmological constant so $w_{\rm de}=-1$, does it have a different, 
constant value, or is it time dependent? For a constant $\wde$,
benchmarks can be given from basic physical principles. To cause
cosmic acceleration a component must have $\wde<-1/3$. Topological
defects such as dominating domain walls have $\wde=-2/3$. If we wish
to distinguish between these fundamental physical mechanisms of a
cosmological constant and domain walls, we need say 
$5\sigma(w_{\rm const})<({}^{-}2/3\,-\,{}^{-}1)$ or $\sigma(w_{\rm const})<0.067$.
This has been achieved by cosmic observations.

There is essentially
no non-finetuned mechanism that has a constant EOS that can cause
acceleration other than topological
defects, so measuring $w$ as a constant (e.g.\ is $w_{\rm const}=-0.95$?)
is no longer a goal. In addition, dynamical $w(a)$ can mimic constant
$w$ through the time averaging inherent over the light propagation of
a cosmic signal, so measuring $w_{\rm const}$, even $w_{\rm const}=-1$
says very little. This is called the mirage of $\Lambda$ \cite{mirage}.

Thus we need a benchmark for $w(a)$. Describing a full function with
just a small number of parameters might appear difficult, but in fact
it is a solved problem. In 2002 \cite{w0wa} demonstrated through study
of the exact solutions of the scalar field equation of motion
(Klein-Gordon equation) that a wide variety of models for dark energy
had their exact predictions for observables (distances, Hubble parameter)
reconstructed to $\sim0.1\%$ accuracy by a remarkably simple form,
\be
w(a)=w_0+w_a(1-a)\ .
\ee
The simplicity of the form should not obscure that this incorporates
the full physics (and does not arise from e.g.\ any artificial Taylor
expansion).

Even models with rapid variations in $w(a)$ have their
observables matched by the $w_0$--$w_a$ parametrization: for example
despite a phase transition the vacuum metamorphosis model \cite{vm,vm2} 
is fit by $(w_0,w_a)=(-1.24,-1.5)$ to 0.55\% in the observables
\cite{1710.02153}; despite oscillations the viable range of
Albrecht-Skordis models \cite{as} are fit by the $w_0$--$w_a$
parametrization to $\lesssim0.1\%$. 

In 2008 \cite{0808.0189} demonstrated that $w_0$--$w_a$ actually
functions as a calibration relation on the dark energy phase EOS phase
space $w$--$w'$, where prime denotes an e-fold derivative $d/d\ln a$.
It cleanly separates different classes of physics behind dark energy.
This important property leads us to well motivated benchmarks for
testing dark energy. The two distinct classes of dark energy dynamics
are freezers and thawers \cite{caldlin}, where the (effective) field either
starts rolling or frozen, respectively, in the high redshift, high Hubble
friction universe. That is, the field evolves toward or away from cosmological
constant behavior respectively.

The class is determined by the steepness of the
(effective) potential. Each class forms a wedge region in the $w$--$w'$
phase space, with the outer edges determined by basic physics: that
the universe was matter dominated at high redshift for the upper boundary
of the thawing region, and that the friction from the Hubble expansion
limits the acceleration of the field for the lower boundary of the
freezing region. The constraints are well understood, and quantified,
in terms of the flow formalism \cite{flow} analogous to that for early
universe inflation.

What is of particular importance for benchmarking is that there is a
gap between the allowed regions, corresponding to the field
acceleration. Only fields that are fine tuned to have zero acceleration
can live more than briefly in this region, so there is a natural
scale to how well $w$--$w'$ needs to be determined to separate which
of the two physics classes describes dark energy. We also emphasize
that this phase space structure has been shown to hold beyond physical
scalar fields, e.g.\ for modified gravity interpreted as an effective
EOS.

The benchmark separation is between the lower bound of the thawing
region, along $w'\approx 1+w$, and the upper bound of the freezing region,
along $w'\approx w(1+w)$. See \cite{paths} for detailed discussion on the
origin of all the boundaries. Thus understanding of the physical
mechanism behind cosmic acceleration, i.e.\ distinction between the
two classes, comes when $\sigma(w')<2(1+w)$, for $w\approx-1$. We
can translate this into a $w_0$--$w_a$ benchmark using the calibration
relation $w_a=-w'(a_\star)/a_\star$ for $a_\star=0.8$ from \cite{0808.0189}. 
Thus our first benchmark criterion is
\be
\sigma(w_a)<2.5\sigma(w_0)\qquad {\rm Benchmark\ Criterion\ 1.}
\ee 

Note that for Stage 4 dark energy experiments we expect roughly
$\sigma(w_0)\approx0.08$--0.1, $\sigma(w_a)\approx0.2$--0.3 including
systematics so we are close to BC1. However, this is a relative
not an absolute criterion; it does not say how accurately either
individual parameter needs to be measured.

We can go further by providing benchmarks for $w_0$ and $w_a$ separately
from either basic physics or motivated models. Recall that
$w'=dw/d\ln a$ measures the scalar field evolution, which is driven
by the steepness of the potential and retarded by Hubble friction.
Within the minimal parameter approach (no extra scales in the problem)
and in the absence of fine tuning, the natural timescale for the scalar field 
to evolve on is the Hubble e-fold scale. That is, $\dot w\sim H^{-1}$.
This translates to $w'\approx1$ or $w_a\approx1$. The other natural
value is no evolution -- that of the cosmological constant. Thus we
want to distinguish between these two natural scales. If we ask for
this important distinction be made at the $5\sigma$ level then
\be
\sigma(w_a)<\frac{1}{5}\times 1=0.2 \qquad {\rm Benchmark\ Criterion\ 2a.}
\ee
This would then give $\sigma(w_0)=0.08$, $\sigma(w_a)=0.2$ as the
criterion, and indeed this is at the Stage 4 level. However, distinction
between evolution at a characteristic e-folding of 0 and of 1 is a
relatively modest goal. While it would give a broad impression of
dark energy, it would not yield critical clues to its specific nature.

Therefore we want to obtain more accuracy, through Stage 5 experiments.
It is more difficult to obtain a general threshold for $w_0$, as it
can approach arbitrarily close to the cosmological constant value of $-1$.
However, an interesting and well motivated model views cosmic acceleration
in a unified manner, across both early universe inflation and late
time acceleration. This ``quintessential inflation'' arises from
$\alpha$-attractor high energy physics, with connections to supergravity
and string theory \cite{1712.09693}. In particular, it predicts a
relation between $w_0$ and the inflation tensor-to-scalar power ratio
$r$. The parameter $r$ is a key goal of next generation cosmic microwave
background experiments detecting primordial gravitational waves through
B-mode polarization.

The $\alpha$-attractor model predicts
$r=12\alpha/N^2$, where $N$ is the number of inflationary e-folds, about
50--60 depending on reheating and model details. Simultaneously
it predicts that after inflation ends and the field freezes again due
to Hubble friction, late time acceleration returns as the universe
expands, the Hubble friction lessens, and the field thaws. The
dark energy then reaches an asymptotic future attractor with
$w_{\rm future}=-1+2/(9\alpha)$. Note the key promise of this theory:
if $\alpha$ is high, we will detect the gravitational wave signature in
the CMB, if $\alpha$ is low we will detect the deviation of the dark
energy EOS from the cosmological constant value! An especially well
motivated value is $\alpha=1$, which corresponds to Starobinsky ($R^2$)
inflation and Higgs inflation. This would give $r\approx4\times 10^{-3}$
and $w_{\rm future}=-0.78$. Today, with $\Omega_{\rm de}\approx 0.7$
the field evolution has not fully reached the future attractor and
$1+w_0\approx0.5(1+w_{\rm future})$, e.g.\ $w_0\approx-0.9$.
If we want to distinguish this from a cosmological constant at $5\sigma$
(just as CMB experiments seek to detect $r$ for this model at $5\sigma$)
then we require 
\be
\sigma(w_0)<\frac{1}{5}\times0.1=0.02 \quad\ {\rm Benchmark\ Criterion\ 2b.} 
\ee

Combining BC2b and BC1 together gives our absolute benchmark criteria for
investigating the nature of dark energy: 
\begin{empheq}[box=\fbox]{align} 
&{\textbf{Expansion\ Benchmark:}}\notag\\ 
\, \\
&\sigma(w_0)=0.02,\ \sigma(w_a)=0.05  \notag 
\end{empheq}


In summary, these two parameters can successfully describe dark energy for
the cosmic expansion history, and physics motivates specific benchmark
criteria for the measurement accuracy required for meaningful understanding.
Stage 4 accuracy will answer some broadbrush questions about dark energy
properties but Stage 5 is required to meet the detailed benchmark criteria.

\section{The Case for $z\approx2-5$} \label{sec:z25}

The $w_0$--$w_a$ parametrization of dark energy fits observations 
with high accuracy, even when the scalar field behavior is more 
extreme, such as phase transitions or dark energy density persisting 
to high redshift (freezing fields and early dark energy). We can, 
however, draw further information about the evolution by looking 
to higher redshifts, $z\gtrsim2$. 

For a late time phase transition, we expect dark energy density 
to vanish at higher redshifts before the phase transition (e.g.\ 
the vacuum metamorphosis model has its phase transition around 
$z\approx1$--2). Conversely, freezing fields and early dark energy 
should have a much higher dark energy density at early times than 
the cosmological constant. At recombination, $z\approx 10^3$, 
the fraction of the critical density in the cosmological constant 
is $\sim 10^{-9}$. However, current constraints on early dark 
energy density are at the $10^{-2.5}$ level. Thus there are more 
than six orders of magnitude of unexplored range. 

While redshifts $z=5$--1000 are fairly inaccessible to precision 
measurement of dark energy, even a cosmological constant contributes 
8\% (1\%) of the critical density at $z=2$ (5), making the dark energy density amenable 
to constraint. If we detected no dark energy density at $z=2$, say, 
this would point us toward phase transition models, while if we 
detected greater than 1\% fractional dark energy density at $z=5$ 
this would favor freezing fields and early dark energy. A $3\sigma$ 
distinction between the cosmological constant behavior and no 
dark energy density (or double the cosmological constant density) 
provides a clear direction for understanding cosmic acceleration 
physics. Thus the benchmark criterion is 
\begin{empheq}[box=\fbox]{align} 
&{\textbf{Density\ Benchmark:}}\notag\\ 
\, \\ 
&\sigma\left(\frac{\rde}{\rho_{\rm crit}}\right)\equiv\sigma(\Omega_{\rm de}(z))<
\frac{1}{3}\,\Omega_{\Lambda}(z)\,, {\rm for\ } z<5\notag
\end{empheq}


\section{Growth and Gravity} \label{sec:gro}

Large scale structure growth can provide incisive constraints 
on dark energy, apart from expansion probes. Within general 
relativity the growth history is essentially determined by the 
expansion history, and so $w_0$--$w_a$ works as well 
for cosmic growth. However, deviations of the growth history behavior 
relative to the expansion history can reveal extra physics beyond 
general relativity. Again, a very successful, single parameter 
measure of aspects of growth separate from that predicted by 
the cosmic expansion was developed in \cite{groexp}. The 
gravitational growth index $\gamma$, given through the 
growth factor behavior 
\be 
\frac{D(a)}{D(a_i)}=e^{\int_{a_i}^a (da/a)\,\om(a)^\gamma}\ , 
\ee 
is accurate to 0.2\% relative to the exact (subhorizon, linear 
perturbation, scale independent) result. For general relativity 
$\gamma\approx0.55$ with only a very mild (and well calibrated) 
dependence on the dark energy EOS. Thus $\gamma$ tests the theory 
of gravity independent of the expansion history. 

Establishing a benchmark for $\gamma$ is difficult to do 
from basic physical principles, since we have a scarcity of 
well motivated theories of gravity apart from general relativity, 
and many theories can approach general relativity arbitrarily 
closely. We do expect that gravity restores to general relativity 
at high redshift, and one can define a phase space evolution 
for the gravitational coupling $\gm$--$\gm'$ \cite{1103.0282}. 
However, it does not separate into distinct regions like the 
$w$--$w'$ phase space. At best we can compare a couple of 
common models: $f(R)$ scalar-tensor gravity and DGP braneworld 
gravity, having $\gamma=0.42$ and $\gamma=0.68$ respectively. 
One could then argue that a $5\sigma$ distinction from general 
relativity requires $\sigma(\gamma)\approx0.026$. In fact, 
due to the phase space evolution, the $\gamma$ value for 
$f(R)$ theories is closer to general relativity's 0.55 in the 
past (roughly half the deviation at $z=1$ on larger scales), 
so for a (admittedly less basic) benchmark criterion we adopt 
\be 
\sigma(\gamma)<\frac{1}{2}\times\frac{1}{5}\times 0.13=0.013\quad\ 
{\rm Growth\ Criterion\ 1.} 
\ee 

While the gravitational growth index is powerful for a 
single parameter, giving a critical alert of growth deviating 
from general relativity, there are two other important avenues 
providing distinction from general relativity. Growth in the linear 
or quasilinear perturbation regime can be scale dependent in 
modified gravity, and light propagation as well as cosmic 
growth can deviate. These can be described (in the subhorizon, 
quasistatic regime) as gravitational couplings entering 
modified Poisson equations for matter and for light 
\cite{bertzukin} in place of Newton's constant. Thus they 
are often called $\gm$ and $\gl$, normalized to unity when 
they equal Newton's constant (there are many other names 
also, e.g.\ $\mu$ and $\Sigma$, but $\gm$ and $\gl$ are the 
most easily understood). 

We can explore deviations from general relativity in a 
model independent manner (i.e.\ without adopting a specific 
theory) by treating the time and scale dependent $\gm(a,k)$ 
and $\gl(a,k)$ as values defined in bins of scale factor $a$ 
and density perturbation wavenumber $k$. Two bins in $k$ and 
three bins in $a$, so six parameters total, have been 
demonstrated to be accurate to 0.1\%--0.3\% rms in the growth 
rate of large scale structure (better on the growth amplitude 
itself) \cite{2208.10508}. In many cases only two bins in $a$ 
are needed for 0.2\% accuracy, giving four parameters. 

As with $\gamma$ it is difficult to develop from first 
principles a benchmark for required accuracy, as gravity 
theories can approach general relativity arbitrarily closely. 
Again looking to the $f(R)$ gravity model, we can seek a 
clear signature of deviation from general relativity in 
scale and time dependence of gravitational coupling, 
adopting 
\bea  
{\rm Growth\ Criterion\ 2.}&\,&\notag\\ 
\sigma(G_{{\rm matter,low\ }k,{\rm high\ }a})&<&0.02\\ 
\sigma(G_{{\rm matter,high\ }k,{\rm high\ }a})&<&0.05\ , 
\eea  
which each correspond to roughly $3\sigma$ constraints 
for a canonical $f(R)$ model. Here low (high) $k$ corresponds 
to $k=0.055$ ($0.125$) $h$/Mpc and high $a$ is $a\approx0.75$ 
($z\approx0.35$). One can use similar criteria 
on $\gl$. 

Another important aspect of modified gravity becomes 
apparent as one approaches the nonlinear perturbation 
regime: a screening mechanism to restore the theory 
to general relativity behavior in regions of high density 
or density gradients (to satisfy solar system tests for 
example). Here the physics becomes complex and model 
dependent enough that it seems worthwhile abandoning 
parametrization and dealing with a full theory, starting 
from its Lagrangian. There are in-between methods such as 
effective field theory, but these do not capture the fully 
nonlinear aspects. At this point we have to choose some 
benchmark models, rather than benchmark principles. This is 
difficult as few to no models are regarded as completely 
compelling by the community as a whole. 

We therefore consider two illustrative models, with the 
aim of keeping them as simple as possible while showcasing 
essential elements of physics. One is $f(R)$ gravity, but 
it is important to recognize that stating this by itself is 
insufficient: one must adopt a particular functional form 
for $f(R)$, with particular parameter values. A common choice 
is the Hu-Sawicki form \cite{husaw}, which has two parameters 
(apart from the density) -- an amplitude $f_{R0}$ and an 
evolution scaling (power law index $n$). We instead adopt 
for simplicity a single parameter exponential $f(R)$ model, 
\be 
f(R)=-b\left(1-e^{-cR/b}\right)\ , 
\ee 
where $b$ is determined by the present matter density 
$\Omega_{m,0}$ and $c$ is the one parameter \cite{0905.2962}. 
Thus our first full gravitational theory benchmark is 
\bea 
{\rm Benchmark\ Gravity\ 1.}&\,&\notag\\ 
{\rm Exponential\ } f(R)\ {\rm with\ }c&=&4  
\eea 
Note that neither exponential nor Hu-Sawicki $f(R)$ 
have a strong theoretical foundation for their 
functional form. The benchmark is simply a point in the 
theory landscape that is sufficiently distinct from general 
relativity to be interesting and currently observationally 
viable. 

One of the drawbacks of using $f(R)$ gravity is 
that it has essentially no effect on light propagation 
different from general relativity. Therefore we adopt a 
second gravity benchmark theory that does affect light 
propagation. 

It is useful to briefly consider the effective 
field theory (EFT) or property function approach \cite{1404.3713}. 
The two key property functions are $\alpha_M$, giving the running 
of the Planck mass arising from coupling of the scalar field 
to the Ricci scalar, and $\alpha_B$, describing the braiding 
between the scalar field kinetic structure and the metric. 
All $f(R)$ gravity theories have $\alpha_B=-\alpha_M$, 
which is also the condition for negligible effect on 
light propagation (there is a residual effect of order $f_{R0}$, 
but this is usually $\lesssim10^{-5}$). This arises from the 
conformal nature of the theory (so null geodesics are unaffected), 
and the coupling to the Ricci scalar can give rise to the 
chameleon mechanism for screening. 

Thus, as a second gravity benchmark theory we will choose 
a theory relying on a different mechanism, called Vainshtein 
screening, which can arise due to the kinetic structure, or 
braiding. Since we also want to have a minimal number of free 
parameters we could simply change the constant of proportionality 
relating $\alpha_B$ to $\alpha_M$ -- for example No Slip 
Gravity has $\alpha_B=-2\alpha_M$ (and then $\gm=\gl$). 
But to emphasize the distinction, and maintain as few 
parameters as possible, we will choose a class of theories 
with standard  
gravitational coupling to the Ricci scalar, so a constant 
Planck mass (i.e.\ $8\pi G_N=m_p^2$) hence $\alpha_M=0$. 
Similarly, the benchmark will have a canonical kinetic 
term (though possibly of opposite sign) with no potential, 
and a cubic Horndeski term only dependent on the field 
kinetic energy $X$, i.e.\ 
$G_3=G_3(X)$, so the action is shift symmetric. 

Kinetic gravity braiding \cite{1008.0048,1103.5360} is 
such a class of theories. The specific form $G_3(X)$ to choose, 
keeping in mind minimal number of parameters, is 
unclear. We know a form $G_3=bX^n$, say, requires 
$n\ge0$ for stability. The most natural choices might 
be $n=0$ -- but this gives $\alpha_B=0$ and no light 
propagation effect; $n=1/2$, i.e.\ $G_3\sim \dot\phi$ -- 
but this can have stability issues; $n=1$, i.e.\ 
$G_3\sim X$ -- but this can have observational 
issues with a wrong sign Sachs-Wolfe effect. 
Also, $G_3\sim \ln X$ 
does not work as it does not restore to 
general relativity at high redshift. 
One could instead work with $\alpha_B$, and 
take a simple form for it such as 
$\alpha_B(a)=A/[1+(4a)^{-3}]$, which has the physically 
expected behaviors of vanishing at high redshift (restoring 
general relativity) and freezing in the de Sitter future. 
However, property 
functions come from a linear level EFT and 
do not fully describe the nonlinear theory (e.g.\ the screening). Thus we leave for future work the determination 
of a one parameter, fully nonlinear theory in the manner 
that $f(R)$ is, that has an effect on light propagation 
and possibly minimal coupling to the Ricci scalar. 
The second full 
gravitational theory benchmark has a placeholder  
\bea 
&\,&{\rm Benchmark\ Gravity\ 2.}\notag\\ 
&\,&{\rm Kinetic\ Gravity\ Braiding\ with\ 1\ parameter?} 
\eea

\section{Achieving Benchmarks} \label{sec:exper}

As mentioned, Stage 4 experiments underway this decade 
will make significant inroads toward quantifying dark energy. 
To achieve the physics derived or motivated benchmarks 
described here will require Stage 5 experiments however. 
While in part such experiments can scale up from Stage 4 
techniques, some different approaches are necessary as well. 

For the Density Benchmark, surveys must extend to $z\approx5$. 
A well tested approach is that of spectroscopic galaxy 
clustering, and plans for a Spectroscopic Stage 5 experiment 
are in design \cite{2209.03585,2209.04322}. This, as well 
as lower redshift techniques such as peculiar velocity surveys,  
high density surveys, and spectrophotometric supernova surveys 
\cite{2005.04325,2203.07291,2105.02204} will also improve 
the Expansion, Growth, and Gravity Benchmarks. 

Significant progress on the challenging benchmark 
criteria for the Expansion Benchmarks (the most motivated 
from physical first principles) likely will rely on more 
than scaling up existing probes. A novel technique coming 
toward maturity is direct measurement of acceleration through 
redshift drift. While other probes involve one or two integrals 
over the dark energy equation of state, redshift drift 
$dz/dt_{\rm obs}=H_0\,(1+z)-H(z)$ is related directly to 
the acceleration $\ddot a$ and hence to $w(a)$. 

Two major advances in the last decade make it viable. On 
the theoretical front it was shown that low redshift 
observations ($z\lesssim0.5$) of the well characterized 
class of emission line galaxies, using {\it differential\/} 
motion of bright forbidden transition doublet lines, was 
optimal. Furthermore, in combination with CMB observations 
redshift drift could achieve the Expansion Benchmark with five 
measurements of 1\% accuracy from $z=0.1$--0.5 \cite{1402.6614}. 
On the experimental front, interferometers 
in tandem with spectrographs have now used differential measurements 
to demonstrate control of systematics $1000\times$ better 
than conventional spectrographs 
\cite{erskine1,erskine2,erskineJatis}. Exoplanet searches 
are driving similar requirements and the technique has been 
demonstrated on the Keck Planet Finder. Improvements 
are being tested currently using Fabry-Perot interferometry 
to further reduce systematics. Most impressively, the 
redshift drift probe simply adds onto existing infrastructure: 
a suitcase size interferometer is used in conjunction with 
an existing spectrograph at a 10 meter telescope (though 
larger telescopes give further S/N improvement). 

Between a cosmic redshift drift experiment and extensions of 
existing probes, a Stage 5 program can achieve the 
necessary benchmarks to answer fundamental physics questions 
about dark energy.

\section{Conclusions} 

While we do not have clear clues as to the physical 
origin of cosmic acceleration and dark energy, we 
do know the questions to ask to obtain insights: 
\begin{enumerate} 
\item There are two natural e-fold timescales for dark 
energy dynamics: 0 and $\mathcal{O}(1)$ -- can we 
distinguish these observationally at $5\sigma$? 
\item There are two natural phase space regions for 
dark energy dynamics: freezing fields moving toward 
cosmological constant behavior and thawing fields moving 
away from it -- can we distinguish these observationally 
at $2\sigma$ (comparing mean of one region to another)? 
\item If cosmic acceleration from inflation is connected 
to late time acceleration through an $\alpha$-attractor 
origin, can we detect its deviation from a cosmological 
constant at $5\sigma$, matching the $5\sigma$ 
detection of primordial gravitational waves in the 
CMB for $\alpha=1$ Starobinsky/Higgs inflation? 
\end{enumerate} 

For further dark energy dynamics we must look beyond 
the region of its domination but where it is still 
observationally relevant, at $z\approx2$--5. Since 
dark energy is allowed to either vanish at high redshift, 
or contribute more than a million times more energy 
density than the cosmological constant by  
recombination, we should measure the dark energy 
density to allow at least a $3\sigma$ determination 
of whether dark energy dies off or significantly 
exceeds $\rho_\Lambda$ out to $z\approx5$. 

Beyond the Expansion Benchmarks and Density Benchmark, 
we must test whether the cosmic growth history is governed 
by the cosmic expansion history in the manner dictated by 
general relativity. In the first instance this can be 
accomplished by separating expansion from growth through 
the gravitational growth index $\gamma$. The next 
key element of understanding comes from the time and 
space dependence of the gravitational coupling, tested 
both for matter and for light. Finally, to access the 
full predictions of a gravity theory into the nonlinear 
regime we adopt two specific benchmark models of 
gravity, using only a single parameter each but with 
a rich array of observational implications. 

All these goals are achievable, starting in the 
next decade and within the 20 year vision. The 
Stage 5 program to answer these fundamental physics 
questions on the nature of dark energy merges 
extensions of current Stage 4 probes plus the highly 
complementary and novel probe of cosmic redshift drift. 
The redshift drift technique is rapidly maturing, 
with strong advances in systematics control (due in 
part to synergy with exoplanet search goals) and 
survey design. 

High energy physics can advance significantly toward 
the ``vitally important'', ``most profound'', and 
``comprehensive fundamental physical'' nature of 
dark energy.

\acknowledgments 

I thank Hitoshi Murayama for interesting discussions 
about dark energy benchmarks. This work 
was supported in part by the U.S.\ Department of Energy, 
Office of Science, Office of High Energy Physics, under 
contract no.\ DE-AC02-05CH11231.


\end{document}